\documentclass{article}

\usepackage{arxiv}

\usepackage[utf8]{inputenc} 
\usepackage[T1]{fontenc}    
\usepackage{hyperref}       
\usepackage{url}            
\usepackage{booktabs}       
\usepackage{amsfonts}       
\usepackage{nicefrac}       
\usepackage{microtype}      
\usepackage{lipsum}

\usepackage{fancyvrb}

\usepackage[nodayofweek]{datetime}

\usepackage{listings}

\usepackage{subfig}

\usepackage{graphicx}

\usepackage{xcolor}

\definecolor{graph_node_entity}{HTML}{66c2a5} 
\definecolor{graph_node_agent}{HTML}{fc8d62} 

\definecolor{graph_edge_contributor}{HTML}{e41a1c} 
\definecolor{graph_edge_team}{HTML}{377eb8} 

\newcommand{\ProvEntity}{$\vcenter{\hbox{\includegraphics[scale=0.12]{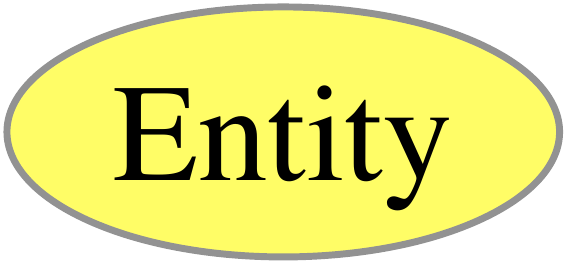}}}$}
\newcommand{\ProvActivity}{$\vcenter{\hbox{\includegraphics[scale=0.12]{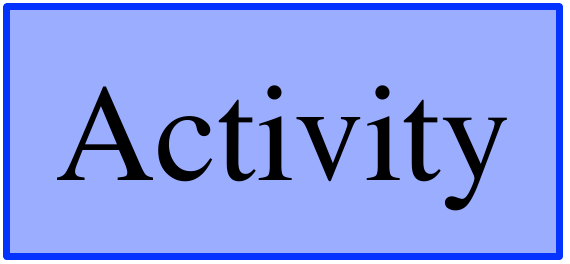}}}$}
\newcommand{\ProvAgent}{$\vcenter{\hbox{\includegraphics[scale=0.12]{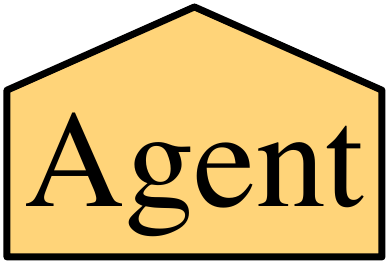}}}$}

\makeatletter
\def\PYG@reset{\let\PYG@it=\relax \let\PYG@bf=\relax%
    \let\PYG@ul=\relax \let\PYG@tc=\relax%
    \let\PYG@bc=\relax \let\PYG@ff=\relax}
\def\PYG@tok#1{\csname PYG@tok@#1\endcsname}
\def\PYG@toks#1+{\ifx\relax#1\empty\else%
    \PYG@tok{#1}\expandafter\PYG@toks\fi}
\def\PYG@do#1{\PYG@bc{\PYG@tc{\PYG@ul{%
    \PYG@it{\PYG@bf{\PYG@ff{#1}}}}}}}
\def\PYG#1#2{\PYG@reset\PYG@toks#1+\relax+\PYG@do{#2}}

\expandafter\def\csname PYG@tok@w\endcsname{\def\PYG@tc##1{\textcolor[rgb]{0.73,0.73,0.73}{##1}}}
\expandafter\def\csname PYG@tok@c\endcsname{\let\PYG@it=\textit\def\PYG@tc##1{\textcolor[rgb]{0.25,0.50,0.50}{##1}}}
\expandafter\def\csname PYG@tok@cp\endcsname{\def\PYG@tc##1{\textcolor[rgb]{0.74,0.48,0.00}{##1}}}
\expandafter\def\csname PYG@tok@k\endcsname{\let\PYG@bf=\textbf\def\PYG@tc##1{\textcolor[rgb]{0.00,0.50,0.00}{##1}}}
\expandafter\def\csname PYG@tok@kp\endcsname{\def\PYG@tc##1{\textcolor[rgb]{0.00,0.50,0.00}{##1}}}
\expandafter\def\csname PYG@tok@kt\endcsname{\def\PYG@tc##1{\textcolor[rgb]{0.69,0.00,0.25}{##1}}}
\expandafter\def\csname PYG@tok@o\endcsname{\def\PYG@tc##1{\textcolor[rgb]{0.40,0.40,0.40}{##1}}}
\expandafter\def\csname PYG@tok@ow\endcsname{\let\PYG@bf=\textbf\def\PYG@tc##1{\textcolor[rgb]{0.67,0.13,1.00}{##1}}}
\expandafter\def\csname PYG@tok@nb\endcsname{\def\PYG@tc##1{\textcolor[rgb]{0.00,0.50,0.00}{##1}}}
\expandafter\def\csname PYG@tok@nf\endcsname{\def\PYG@tc##1{\textcolor[rgb]{0.00,0.00,1.00}{##1}}}
\expandafter\def\csname PYG@tok@nc\endcsname{\let\PYG@bf=\textbf\def\PYG@tc##1{\textcolor[rgb]{0.00,0.00,1.00}{##1}}}
\expandafter\def\csname PYG@tok@nn\endcsname{\let\PYG@bf=\textbf\def\PYG@tc##1{\textcolor[rgb]{0.00,0.00,1.00}{##1}}}
\expandafter\def\csname PYG@tok@ne\endcsname{\let\PYG@bf=\textbf\def\PYG@tc##1{\textcolor[rgb]{0.82,0.25,0.23}{##1}}}
\expandafter\def\csname PYG@tok@nv\endcsname{\def\PYG@tc##1{\textcolor[rgb]{0.10,0.09,0.49}{##1}}}
\expandafter\def\csname PYG@tok@no\endcsname{\def\PYG@tc##1{\textcolor[rgb]{0.53,0.00,0.00}{##1}}}
\expandafter\def\csname PYG@tok@nl\endcsname{\def\PYG@tc##1{\textcolor[rgb]{0.63,0.63,0.00}{##1}}}
\expandafter\def\csname PYG@tok@ni\endcsname{\let\PYG@bf=\textbf\def\PYG@tc##1{\textcolor[rgb]{0.60,0.60,0.60}{##1}}}
\expandafter\def\csname PYG@tok@na\endcsname{\def\PYG@tc##1{\textcolor[rgb]{0.49,0.56,0.16}{##1}}}
\expandafter\def\csname PYG@tok@nt\endcsname{\let\PYG@bf=\textbf\def\PYG@tc##1{\textcolor[rgb]{0.00,0.50,0.00}{##1}}}
\expandafter\def\csname PYG@tok@nd\endcsname{\def\PYG@tc##1{\textcolor[rgb]{0.67,0.13,1.00}{##1}}}
\expandafter\def\csname PYG@tok@s\endcsname{\def\PYG@tc##1{\textcolor[rgb]{0.73,0.13,0.13}{##1}}}
\expandafter\def\csname PYG@tok@sd\endcsname{\let\PYG@it=\textit\def\PYG@tc##1{\textcolor[rgb]{0.73,0.13,0.13}{##1}}}
\expandafter\def\csname PYG@tok@si\endcsname{\let\PYG@bf=\textbf\def\PYG@tc##1{\textcolor[rgb]{0.73,0.40,0.53}{##1}}}
\expandafter\def\csname PYG@tok@se\endcsname{\let\PYG@bf=\textbf\def\PYG@tc##1{\textcolor[rgb]{0.73,0.40,0.13}{##1}}}
\expandafter\def\csname PYG@tok@sr\endcsname{\def\PYG@tc##1{\textcolor[rgb]{0.73,0.40,0.53}{##1}}}
\expandafter\def\csname PYG@tok@ss\endcsname{\def\PYG@tc##1{\textcolor[rgb]{0.10,0.09,0.49}{##1}}}
\expandafter\def\csname PYG@tok@sx\endcsname{\def\PYG@tc##1{\textcolor[rgb]{0.00,0.50,0.00}{##1}}}
\expandafter\def\csname PYG@tok@m\endcsname{\def\PYG@tc##1{\textcolor[rgb]{0.40,0.40,0.40}{##1}}}
\expandafter\def\csname PYG@tok@gh\endcsname{\let\PYG@bf=\textbf\def\PYG@tc##1{\textcolor[rgb]{0.00,0.00,0.50}{##1}}}
\expandafter\def\csname PYG@tok@gu\endcsname{\let\PYG@bf=\textbf\def\PYG@tc##1{\textcolor[rgb]{0.50,0.00,0.50}{##1}}}
\expandafter\def\csname PYG@tok@gd\endcsname{\def\PYG@tc##1{\textcolor[rgb]{0.63,0.00,0.00}{##1}}}
\expandafter\def\csname PYG@tok@gi\endcsname{\def\PYG@tc##1{\textcolor[rgb]{0.00,0.63,0.00}{##1}}}
\expandafter\def\csname PYG@tok@gr\endcsname{\def\PYG@tc##1{\textcolor[rgb]{1.00,0.00,0.00}{##1}}}
\expandafter\def\csname PYG@tok@ge\endcsname{\let\PYG@it=\textit}
\expandafter\def\csname PYG@tok@gs\endcsname{\let\PYG@bf=\textbf}
\expandafter\def\csname PYG@tok@gp\endcsname{\let\PYG@bf=\textbf\def\PYG@tc##1{\textcolor[rgb]{0.00,0.00,0.50}{##1}}}
\expandafter\def\csname PYG@tok@go\endcsname{\def\PYG@tc##1{\textcolor[rgb]{0.53,0.53,0.53}{##1}}}
\expandafter\def\csname PYG@tok@gt\endcsname{\def\PYG@tc##1{\textcolor[rgb]{0.00,0.27,0.87}{##1}}}
\expandafter\def\csname PYG@tok@err\endcsname{\def\PYG@bc##1{\setlength{\fboxsep}{0pt}\fcolorbox[rgb]{1.00,0.00,0.00}{1,1,1}{\strut ##1}}}
\expandafter\def\csname PYG@tok@kc\endcsname{\let\PYG@bf=\textbf\def\PYG@tc##1{\textcolor[rgb]{0.00,0.50,0.00}{##1}}}
\expandafter\def\csname PYG@tok@kd\endcsname{\let\PYG@bf=\textbf\def\PYG@tc##1{\textcolor[rgb]{0.00,0.50,0.00}{##1}}}
\expandafter\def\csname PYG@tok@kn\endcsname{\let\PYG@bf=\textbf\def\PYG@tc##1{\textcolor[rgb]{0.00,0.50,0.00}{##1}}}
\expandafter\def\csname PYG@tok@kr\endcsname{\let\PYG@bf=\textbf\def\PYG@tc##1{\textcolor[rgb]{0.00,0.50,0.00}{##1}}}
\expandafter\def\csname PYG@tok@bp\endcsname{\def\PYG@tc##1{\textcolor[rgb]{0.00,0.50,0.00}{##1}}}
\expandafter\def\csname PYG@tok@fm\endcsname{\def\PYG@tc##1{\textcolor[rgb]{0.00,0.00,1.00}{##1}}}
\expandafter\def\csname PYG@tok@vc\endcsname{\def\PYG@tc##1{\textcolor[rgb]{0.10,0.09,0.49}{##1}}}
\expandafter\def\csname PYG@tok@vg\endcsname{\def\PYG@tc##1{\textcolor[rgb]{0.10,0.09,0.49}{##1}}}
\expandafter\def\csname PYG@tok@vi\endcsname{\def\PYG@tc##1{\textcolor[rgb]{0.10,0.09,0.49}{##1}}}
\expandafter\def\csname PYG@tok@vm\endcsname{\def\PYG@tc##1{\textcolor[rgb]{0.10,0.09,0.49}{##1}}}
\expandafter\def\csname PYG@tok@sa\endcsname{\def\PYG@tc##1{\textcolor[rgb]{0.73,0.13,0.13}{##1}}}
\expandafter\def\csname PYG@tok@sb\endcsname{\def\PYG@tc##1{\textcolor[rgb]{0.73,0.13,0.13}{##1}}}
\expandafter\def\csname PYG@tok@sc\endcsname{\def\PYG@tc##1{\textcolor[rgb]{0.73,0.13,0.13}{##1}}}
\expandafter\def\csname PYG@tok@dl\endcsname{\def\PYG@tc##1{\textcolor[rgb]{0.73,0.13,0.13}{##1}}}
\expandafter\def\csname PYG@tok@s2\endcsname{\def\PYG@tc##1{\textcolor[rgb]{0.73,0.13,0.13}{##1}}}
\expandafter\def\csname PYG@tok@sh\endcsname{\def\PYG@tc##1{\textcolor[rgb]{0.73,0.13,0.13}{##1}}}
\expandafter\def\csname PYG@tok@s1\endcsname{\def\PYG@tc##1{\textcolor[rgb]{0.73,0.13,0.13}{##1}}}
\expandafter\def\csname PYG@tok@mb\endcsname{\def\PYG@tc##1{\textcolor[rgb]{0.40,0.40,0.40}{##1}}}
\expandafter\def\csname PYG@tok@mf\endcsname{\def\PYG@tc##1{\textcolor[rgb]{0.40,0.40,0.40}{##1}}}
\expandafter\def\csname PYG@tok@mh\endcsname{\def\PYG@tc##1{\textcolor[rgb]{0.40,0.40,0.40}{##1}}}
\expandafter\def\csname PYG@tok@mi\endcsname{\def\PYG@tc##1{\textcolor[rgb]{0.40,0.40,0.40}{##1}}}
\expandafter\def\csname PYG@tok@il\endcsname{\def\PYG@tc##1{\textcolor[rgb]{0.40,0.40,0.40}{##1}}}
\expandafter\def\csname PYG@tok@mo\endcsname{\def\PYG@tc##1{\textcolor[rgb]{0.40,0.40,0.40}{##1}}}
\expandafter\def\csname PYG@tok@ch\endcsname{\let\PYG@it=\textit\def\PYG@tc##1{\textcolor[rgb]{0.25,0.50,0.50}{##1}}}
\expandafter\def\csname PYG@tok@cm\endcsname{\let\PYG@it=\textit\def\PYG@tc##1{\textcolor[rgb]{0.25,0.50,0.50}{##1}}}
\expandafter\def\csname PYG@tok@cpf\endcsname{\let\PYG@it=\textit\def\PYG@tc##1{\textcolor[rgb]{0.25,0.50,0.50}{##1}}}
\expandafter\def\csname PYG@tok@c1\endcsname{\let\PYG@it=\textit\def\PYG@tc##1{\textcolor[rgb]{0.25,0.50,0.50}{##1}}}
\expandafter\def\csname PYG@tok@cs\endcsname{\let\PYG@it=\textit\def\PYG@tc##1{\textcolor[rgb]{0.25,0.50,0.50}{##1}}}

\def\PYGZus{\char`\_}
\def\PYGZob{\char`\{}
\def\PYGZcb{\char`\}}

\def\PYGZlt{\char`\<}
\def\PYGZgt{\char`\>}

\def\PYGZhy{\char`\-}
\def\PYGZsq{\char`\'}

\makeatother

\makeatletter
\def\PYGabap@reset{\let\PYGabap@it=\relax \let\PYGabap@bf=\relax%
    \let\PYGabap@ul=\relax \let\PYGabap@tc=\relax%
    \let\PYGabap@bc=\relax \let\PYGabap@ff=\relax}
\def\PYGabap@tok#1{\csname PYGabap@tok@#1\endcsname}
\def\PYGabap@toks#1+{\ifx\relax#1\empty\else%
    \PYGabap@tok{#1}\expandafter\PYGabap@toks\fi}
\def\PYGabap@do#1{\PYGabap@bc{\PYGabap@tc{\PYGabap@ul{%
    \PYGabap@it{\PYGabap@bf{\PYGabap@ff{#1}}}}}}}
\def\PYGabap#1#2{\PYGabap@reset\PYGabap@toks#1+\relax+\PYGabap@do{#2}}

\expandafter\def\csname PYGabap@tok@c\endcsname{\let\PYGabap@it=\textit\def\PYGabap@tc##1{\textcolor[rgb]{0.53,0.53,0.53}{##1}}}
\expandafter\def\csname PYGabap@tok@cs\endcsname{\let\PYGabap@it=\textit\def\PYGabap@tc##1{\textcolor[rgb]{0.53,0.53,0.53}{##1}}}
\expandafter\def\csname PYGabap@tok@k\endcsname{\def\PYGabap@tc##1{\textcolor[rgb]{0.00,0.00,1.00}{##1}}}
\expandafter\def\csname PYGabap@tok@ow\endcsname{\def\PYGabap@tc##1{\textcolor[rgb]{0.00,0.00,1.00}{##1}}}
\expandafter\def\csname PYGabap@tok@n\endcsname{\def\PYGabap@tc##1{\textcolor[rgb]{0.00,0.00,0.00}{##1}}}
\expandafter\def\csname PYGabap@tok@m\endcsname{\def\PYGabap@tc##1{\textcolor[rgb]{0.20,0.67,1.00}{##1}}}
\expandafter\def\csname PYGabap@tok@s\endcsname{\def\PYGabap@tc##1{\textcolor[rgb]{0.33,0.67,0.13}{##1}}}
\expandafter\def\csname PYGabap@tok@err\endcsname{\def\PYGabap@tc##1{\textcolor[rgb]{1.00,0.00,0.00}{##1}}}
\expandafter\def\csname PYGabap@tok@kc\endcsname{\def\PYGabap@tc##1{\textcolor[rgb]{0.00,0.00,1.00}{##1}}}
\expandafter\def\csname PYGabap@tok@kd\endcsname{\def\PYGabap@tc##1{\textcolor[rgb]{0.00,0.00,1.00}{##1}}}
\expandafter\def\csname PYGabap@tok@kn\endcsname{\def\PYGabap@tc##1{\textcolor[rgb]{0.00,0.00,1.00}{##1}}}
\expandafter\def\csname PYGabap@tok@kp\endcsname{\def\PYGabap@tc##1{\textcolor[rgb]{0.00,0.00,1.00}{##1}}}
\expandafter\def\csname PYGabap@tok@kr\endcsname{\def\PYGabap@tc##1{\textcolor[rgb]{0.00,0.00,1.00}{##1}}}
\expandafter\def\csname PYGabap@tok@kt\endcsname{\def\PYGabap@tc##1{\textcolor[rgb]{0.00,0.00,1.00}{##1}}}
\expandafter\def\csname PYGabap@tok@na\endcsname{\def\PYGabap@tc##1{\textcolor[rgb]{0.00,0.00,0.00}{##1}}}
\expandafter\def\csname PYGabap@tok@nb\endcsname{\def\PYGabap@tc##1{\textcolor[rgb]{0.00,0.00,0.00}{##1}}}
\expandafter\def\csname PYGabap@tok@bp\endcsname{\def\PYGabap@tc##1{\textcolor[rgb]{0.00,0.00,0.00}{##1}}}
\expandafter\def\csname PYGabap@tok@nc\endcsname{\def\PYGabap@tc##1{\textcolor[rgb]{0.00,0.00,0.00}{##1}}}
\expandafter\def\csname PYGabap@tok@no\endcsname{\def\PYGabap@tc##1{\textcolor[rgb]{0.00,0.00,0.00}{##1}}}
\expandafter\def\csname PYGabap@tok@nd\endcsname{\def\PYGabap@tc##1{\textcolor[rgb]{0.00,0.00,0.00}{##1}}}
\expandafter\def\csname PYGabap@tok@ni\endcsname{\def\PYGabap@tc##1{\textcolor[rgb]{0.00,0.00,0.00}{##1}}}
\expandafter\def\csname PYGabap@tok@ne\endcsname{\def\PYGabap@tc##1{\textcolor[rgb]{0.00,0.00,0.00}{##1}}}
\expandafter\def\csname PYGabap@tok@nf\endcsname{\def\PYGabap@tc##1{\textcolor[rgb]{0.00,0.00,0.00}{##1}}}
\expandafter\def\csname PYGabap@tok@fm\endcsname{\def\PYGabap@tc##1{\textcolor[rgb]{0.00,0.00,0.00}{##1}}}
\expandafter\def\csname PYGabap@tok@py\endcsname{\def\PYGabap@tc##1{\textcolor[rgb]{0.00,0.00,0.00}{##1}}}
\expandafter\def\csname PYGabap@tok@nl\endcsname{\def\PYGabap@tc##1{\textcolor[rgb]{0.00,0.00,0.00}{##1}}}
\expandafter\def\csname PYGabap@tok@nn\endcsname{\def\PYGabap@tc##1{\textcolor[rgb]{0.00,0.00,0.00}{##1}}}
\expandafter\def\csname PYGabap@tok@nx\endcsname{\def\PYGabap@tc##1{\textcolor[rgb]{0.00,0.00,0.00}{##1}}}
\expandafter\def\csname PYGabap@tok@nt\endcsname{\def\PYGabap@tc##1{\textcolor[rgb]{0.00,0.00,0.00}{##1}}}
\expandafter\def\csname PYGabap@tok@nv\endcsname{\def\PYGabap@tc##1{\textcolor[rgb]{0.00,0.00,0.00}{##1}}}
\expandafter\def\csname PYGabap@tok@vc\endcsname{\def\PYGabap@tc##1{\textcolor[rgb]{0.00,0.00,0.00}{##1}}}
\expandafter\def\csname PYGabap@tok@vg\endcsname{\def\PYGabap@tc##1{\textcolor[rgb]{0.00,0.00,0.00}{##1}}}
\expandafter\def\csname PYGabap@tok@vi\endcsname{\def\PYGabap@tc##1{\textcolor[rgb]{0.00,0.00,0.00}{##1}}}
\expandafter\def\csname PYGabap@tok@vm\endcsname{\def\PYGabap@tc##1{\textcolor[rgb]{0.00,0.00,0.00}{##1}}}
\expandafter\def\csname PYGabap@tok@sa\endcsname{\def\PYGabap@tc##1{\textcolor[rgb]{0.33,0.67,0.13}{##1}}}
\expandafter\def\csname PYGabap@tok@sb\endcsname{\def\PYGabap@tc##1{\textcolor[rgb]{0.33,0.67,0.13}{##1}}}
\expandafter\def\csname PYGabap@tok@sc\endcsname{\def\PYGabap@tc##1{\textcolor[rgb]{0.33,0.67,0.13}{##1}}}
\expandafter\def\csname PYGabap@tok@dl\endcsname{\def\PYGabap@tc##1{\textcolor[rgb]{0.33,0.67,0.13}{##1}}}
\expandafter\def\csname PYGabap@tok@sd\endcsname{\def\PYGabap@tc##1{\textcolor[rgb]{0.33,0.67,0.13}{##1}}}
\expandafter\def\csname PYGabap@tok@s2\endcsname{\def\PYGabap@tc##1{\textcolor[rgb]{0.33,0.67,0.13}{##1}}}
\expandafter\def\csname PYGabap@tok@se\endcsname{\def\PYGabap@tc##1{\textcolor[rgb]{0.33,0.67,0.13}{##1}}}
\expandafter\def\csname PYGabap@tok@sh\endcsname{\def\PYGabap@tc##1{\textcolor[rgb]{0.33,0.67,0.13}{##1}}}
\expandafter\def\csname PYGabap@tok@si\endcsname{\def\PYGabap@tc##1{\textcolor[rgb]{0.33,0.67,0.13}{##1}}}
\expandafter\def\csname PYGabap@tok@sx\endcsname{\def\PYGabap@tc##1{\textcolor[rgb]{0.33,0.67,0.13}{##1}}}
\expandafter\def\csname PYGabap@tok@sr\endcsname{\def\PYGabap@tc##1{\textcolor[rgb]{0.33,0.67,0.13}{##1}}}
\expandafter\def\csname PYGabap@tok@s1\endcsname{\def\PYGabap@tc##1{\textcolor[rgb]{0.33,0.67,0.13}{##1}}}
\expandafter\def\csname PYGabap@tok@ss\endcsname{\def\PYGabap@tc##1{\textcolor[rgb]{0.33,0.67,0.13}{##1}}}
\expandafter\def\csname PYGabap@tok@mb\endcsname{\def\PYGabap@tc##1{\textcolor[rgb]{0.20,0.67,1.00}{##1}}}
\expandafter\def\csname PYGabap@tok@mf\endcsname{\def\PYGabap@tc##1{\textcolor[rgb]{0.20,0.67,1.00}{##1}}}
\expandafter\def\csname PYGabap@tok@mh\endcsname{\def\PYGabap@tc##1{\textcolor[rgb]{0.20,0.67,1.00}{##1}}}
\expandafter\def\csname PYGabap@tok@mi\endcsname{\def\PYGabap@tc##1{\textcolor[rgb]{0.20,0.67,1.00}{##1}}}
\expandafter\def\csname PYGabap@tok@il\endcsname{\def\PYGabap@tc##1{\textcolor[rgb]{0.20,0.67,1.00}{##1}}}
\expandafter\def\csname PYGabap@tok@mo\endcsname{\def\PYGabap@tc##1{\textcolor[rgb]{0.20,0.67,1.00}{##1}}}
\expandafter\def\csname PYGabap@tok@ch\endcsname{\let\PYGabap@it=\textit\def\PYGabap@tc##1{\textcolor[rgb]{0.53,0.53,0.53}{##1}}}
\expandafter\def\csname PYGabap@tok@cm\endcsname{\let\PYGabap@it=\textit\def\PYGabap@tc##1{\textcolor[rgb]{0.53,0.53,0.53}{##1}}}
\expandafter\def\csname PYGabap@tok@cp\endcsname{\let\PYGabap@it=\textit\def\PYGabap@tc##1{\textcolor[rgb]{0.53,0.53,0.53}{##1}}}
\expandafter\def\csname PYGabap@tok@cpf\endcsname{\let\PYGabap@it=\textit\def\PYGabap@tc##1{\textcolor[rgb]{0.53,0.53,0.53}{##1}}}
\expandafter\def\csname PYGabap@tok@c1\endcsname{\let\PYGabap@it=\textit\def\PYGabap@tc##1{\textcolor[rgb]{0.53,0.53,0.53}{##1}}}

\makeatother

\title{Visualization of Contributions to Open-Source Projects}

\author{
  Andreas Schreiber\thanks{\url{https://www.dlr.de/sc/ivs}} \\
  Intelligent and Distributed Systems\\
  Institute for Software Technology\\
  German Aerospace Center (DLR)\\
  51147 K{\"o}ln, Germany \\
  \texttt{andreas.schreiber@dlr.de} \\
}

\begin{document}
\maketitle

\begin{abstract}
We want to analyze visually, to what extend team members and external developers contribute to open-source projects. This gives a high-level impression about collaboration in that projects. We achieve this by recording provenance of the development process and use graph drawing on the resulting provenance graph. Our graph drawings show, which developers are jointly changed the same files---and to what extend---which we show at Germany's COVID-19 exposure notification app 'Corona-Warn-App'. 
\end{abstract}

\keywords{graph visualization \and software visualization \and provenance \and open source}

\section{Introduction}
In open-source projects, team composition and development process is transparent and traceable, which is one of the advantages of the open-source model~\cite{MidhaPalvia2012}. Understanding of patterns and characteristics of open-source projects, where---sometime many---developers with different roles~\cite{WangFengWangEtAl2020} work together, is an important research question; especially for projects with high public interests.

The COVID-19 pandemic~\cite{ChenAllotLu2020} raises challenges for scientists of many disciplines. Computer scientists and software developers help to fight the pandemic with software systems, which must be developed under time pressure~\cite{CarrollConboy2020}. For example, apps for mobile devices that support \emph{contact tracing} of infected persons are useful to identify local COVID-19 hot-spots and find other persons, who are potentially infected, too. We focus on Germany's exposure notification app \textit{Corona-Warn-App} (CWA; see Section~\ref{sec:example_cwa})\@.

For the CWA, we want to analyze and see visually, to what extend team members and external contributors contributed to the various sub-projects of CWA on GitHub. 

Our method is to record the \emph{provenance of software development processes}~\cite{MoreauGrothMilesEtAl2008,WendelKundeSchreiber2010} and store it according to a standard provenance data model. Technically, we do repository mining to extract provenance and store it as a labeled property graph in graph databases. We query the graph for information to answer our research questions directly or for parts of the graph to visualize it with ``standard'' graph drawing.

We describe the contributions and the emerging results of our works as follows:
\begin{itemize}
    \item A brief description of provenance of software development processes with focus on open-source processes that use the version control system \texttt{git} (Section~\ref{sec:provenance}).
    \item An overview how we draw graphs that visually show contributions by developers with different roles (Section~\ref{sec:graph_visualization}).
    \item As an example towards a user study, we present graph drawings for the Corona-Warn-App (Section~\ref{sec:example_cwa}).
\end{itemize}

\section{Provenance of Software Development Processes}
\label{sec:provenance}

Provenance can be expressed in many formats. We focus on the standard W3C \textsc{PROV}~\cite{MoreauGroth2013}, which defines the provenance data model \textsc{PROV-DM}~\cite{MoreauMissierBelhajjameEtAl2013}. The core structure of \textsc{PROV-DM} relies on the definition of the model class elements \emph{entities}~\ProvEntity, \emph{activities}~\ProvActivity, and \emph{agents}~\ProvAgent~that are involved in producing a piece of data or artifact and on definitions of \emph{relations} to relate these class elements, such as \emph{wasGeneratedBy}, \emph{wasAssociatedWith}, \emph{wasAttributedTo}, and \emph{used}. Each of the class elements and relations can have additional attributes.

\begin{figure}[h]
  \centering
  \includegraphics[width=0.5\columnwidth]{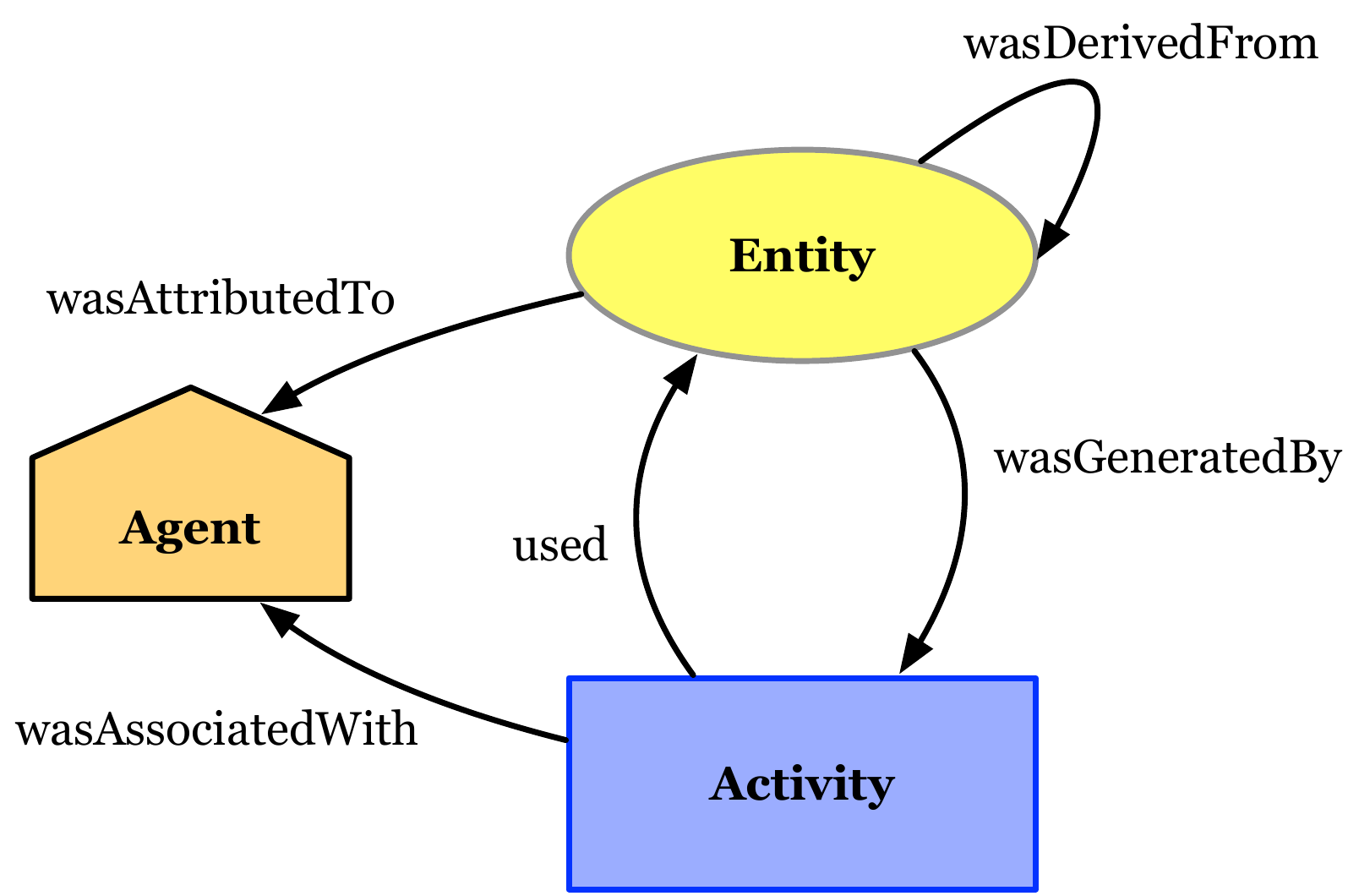}
  \caption{Overview of the PROV model: class elements \emph{entities}, \emph{activities}, and \emph{agents} with relations.}
  \label{fig:prov-overview}
\end{figure}

Provenance of an entity (e.g., a software artefact) is a \emph{directed acyclic graph} (DAG). Since all nodes and edges of this graph have a defined semantics, the provenance graph is a specific \emph{knowledge graph}. The provenance graph can be stored in graph databases as a \emph{labeled property graph}.

\subsection{Provenance for \texttt{git} Repositories}

\begin{figure}[ht]
  \centering
  \includegraphics[width=0.8\columnwidth]{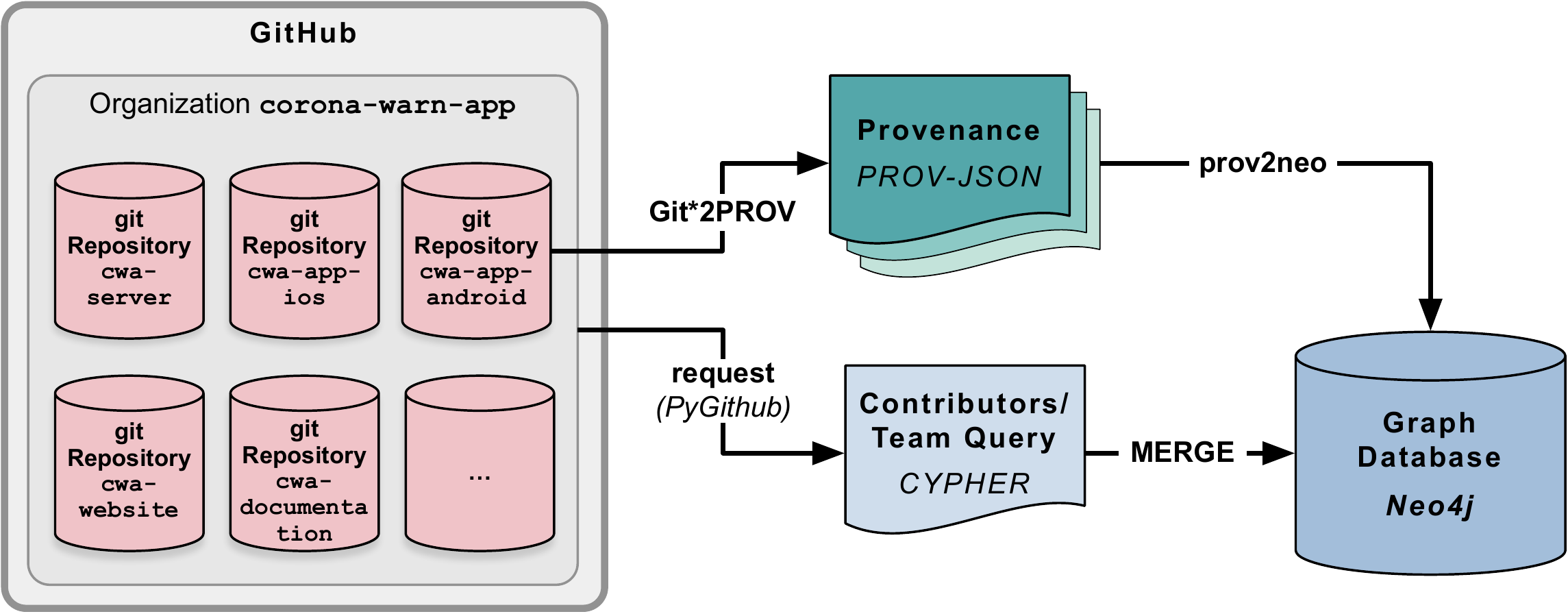}
  \caption{Extracting provenance from git repositories.}
  \label{fig:repository-mining}
\end{figure}

To analyze software development processes, we extract \emph{retrospective provenance}~\cite{McPhillipsBowersBelhajjameEtAl2015} from repositories and store it in a graph database for further analysis (Figure~\ref{fig:repository-mining})~\cite{SchreiberBoer2020}.

To extract provenance from \texttt{git}-based projects we use tools, which crawl the \texttt{git} repositories and additional information, such as issues or pull requests (Git2PROV~\cite{DeNiesMagliacaneVerborghEtAl2013,VerborghMagliacaneSchreiberEtAl2020} and GitHub2PROV~\cite{PackerChapmanCarr2019}). 

\subsection{Using and Analyzing Provenance}

To analyze provenance graphs, many \emph{visual} and \emph{analytical} methods exist---including graph summarization~\cite{Moreau2015,TianHankinsPatel2008}, or visual exploration~\cite{Wattenberg2006} . 

For example, we illustrate querying and using the provenance graph to answer the question: ``\emph{Which files have commits by team members as well as external contributors?}''

We generate a \textsc{Cypher} query, that adds information about contributors roles. We retrieve member information via the GitHub API and store it in Python lists of team members and external contributors, which we insert in a \textsc{Cypher} template.
This \textsc{Cypher} query creates new directed relations between persons~\ProvAgent~ and files~\ProvEntity; for example, the relation for team members is: 
\begin{Verbatim}[commandchars=\\\{\}]
  \PYG{p}{(:}\PYG{n}{Agent}\PYG{p}{)}\PYG{o}{\PYGZhy{}[}\PYG{p}{:}\PYG{n}{CONTRIBUTES\PYGZus{}TO}\PYG{+w}{ }\PYG{p}{\PYGZob{}}\PYG{n}{role}\PYG{p}{:}\PYG{+w}{ }\PYG{err}{\PYGZsq{}}\PYG{n}{team}\PYG{err}{\PYGZsq{}}\PYG{p}{\PYGZcb{}}\PYG{o}{]\PYGZhy{}\PYGZgt{}}\PYG{p}{(:}\PYG{n}{Entity}\PYG{p}{)}
  \end{Verbatim}
  

Then we query for files, where team members and external contributor made changes at any of the files revisions (the query result is exported for visualization (see Section~\ref{sec:graph_visualization}):

\begin{Verbatim}[commandchars=\\\{\}]
  \PYG{k}{MATCH}
  \PYG{+w}{  }\PYG{p}{(}\PYG{n}{team\PYGZus{}member}\PYG{p}{:}\PYG{n}{Agent}\PYG{p}{)}
  \PYG{+w}{  }\PYG{err}{\PYGZhy{}}\PYG{o}{[}\PYG{n}{r1}\PYG{p}{:}\PYG{n}{CONTRIBUTES\PYGZus{}TO}\PYG{+w}{ }\PYG{p}{\PYGZob{}}\PYG{n}{role}\PYG{p}{:}\PYG{+w}{ }\PYG{err}{\PYGZsq{}}\PYG{n}{team}\PYG{err}{\PYGZsq{}}\PYG{p}{\PYGZcb{}}\PYG{o}{]}
  \PYG{+w}{  }\PYG{err}{\PYGZhy{}}\PYG{p}{\PYGZgt{}(}\PYG{n}{file}\PYG{p}{:}\PYG{n}{Entity}\PYG{p}{)}
  \PYG{+w}{  }\PYG{p}{\PYGZlt{}}\PYG{err}{\PYGZhy{}}\PYG{o}{[}\PYG{n}{r2}\PYG{p}{:}\PYG{n}{CONTRIBUTES\PYGZus{}TO}\PYG{+w}{ }\PYG{p}{\PYGZob{}}\PYG{n}{role}\PYG{p}{:}\PYG{+w}{ }\PYG{err}{\PYGZsq{}}\PYG{n}{contributor}\PYG{err}{\PYGZsq{}}\PYG{p}{\PYGZcb{}}\PYG{o}{]}
  \PYG{+w}{  }\PYG{err}{\PYGZhy{}}\PYG{p}{(}\PYG{n}{external\PYGZus{}contributor}\PYG{p}{:}\PYG{n}{Agent}\PYG{p}{)}
  \PYG{k}{RETURN}
  \PYG{+w}{  }\PYG{n}{team\PYGZus{}member}\PYG{p}{,}\PYG{n}{file}\PYG{p}{,}\PYG{n}{external\PYGZus{}contributor}
  \end{Verbatim}

\section{Graph Visualization}
\label{sec:graph_visualization}

We visualize parts of the property graph that is derived from the provenance graph. We use a graph visualization that is \emph{readable} and \emph{faithful}~\cite{NguyenEadesHong2013,NguyenEades2017}.

\begin{figure}[ht]
  \centering
  \includegraphics[width=0.8\columnwidth]{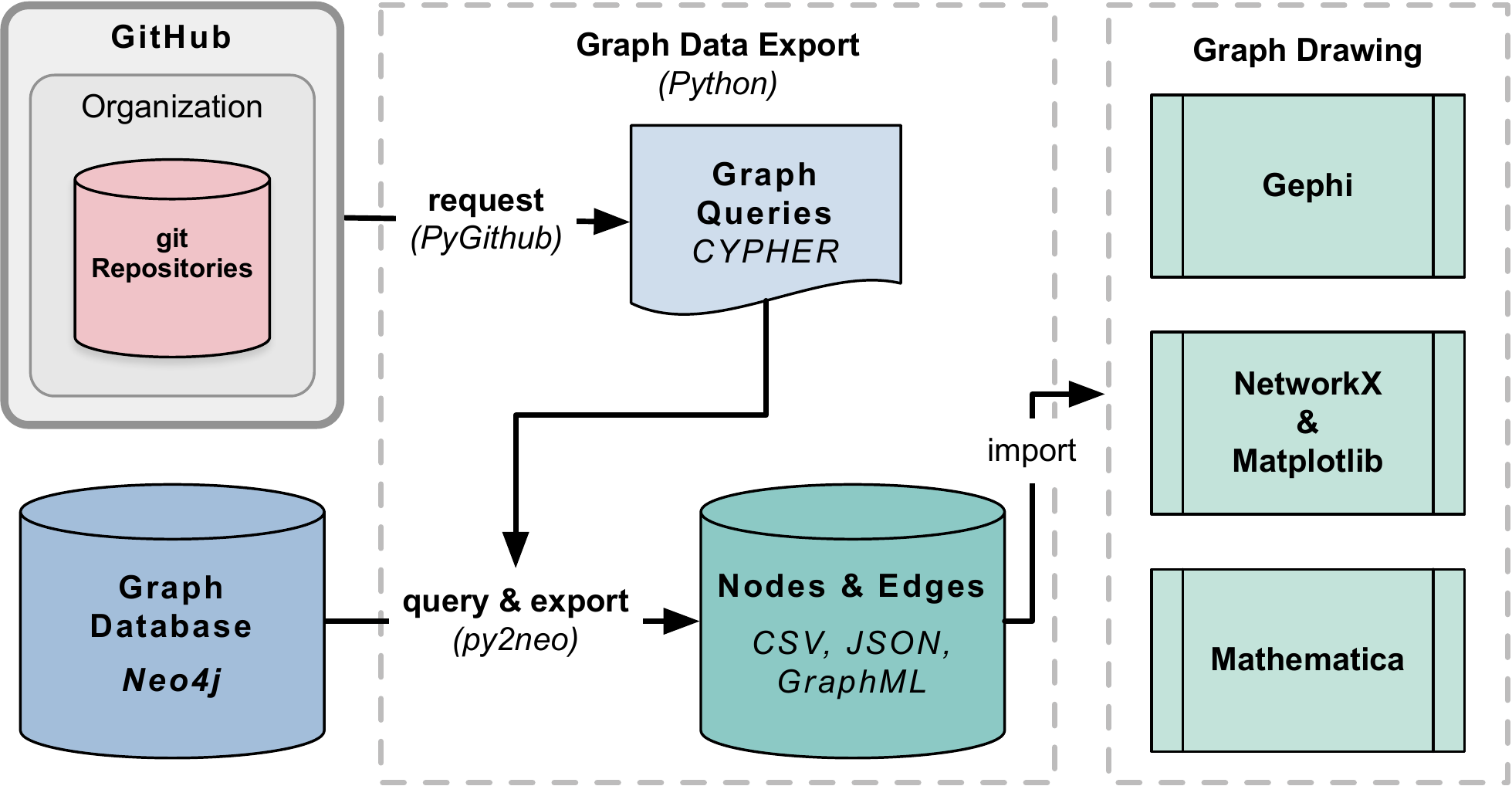}
  \caption{Querying and exporting graph data for visualization; there are multiple choices possible for graph drawing, such as \textsc{Gephi}, Python with \texttt{networkx} and \texttt{matplotlib}, or \textsc{Mathematica}.}
  \label{fig:graph-visualization}
\end{figure}

Using a Python script, we export the relevant nodes and edges from \textsc{Neo4j} and store them in intermediate files; specifically in CSV, JSON, and GraphML files, which we import into graph drawing software (Figure~\ref{fig:graph-visualization}). In the following, we use \textsc{Gephi}~\cite{BastianHeymannJacomy2009} to draw our graphs.

During querying and exporting for visualization, we map the property graph as follows:

\begin{itemize}
    \item PROV elements \emph{entities}~\ProvEntity~ (i.e., files) and \emph{agents}~\ProvAgent~ 
    (i.e., contributors) become graph nodes with two distinct colors.
    \item The relations CONTRIBUTES\_TO become edges, which color depends on the property \emph{role}.
\end{itemize}

For the \emph{coloring}~\cite{KarimKwonParkEtAl2019}, we choose distinct colors from two different qualitative color schemes generated by \textsc{ColorBrewer}~\cite{HarrowerBrewer2003}. Nodes use colors from the \emph{``3-class Set2''} schema\footnote{\url{https://colorbrewer2.org/?type=qualitative&scheme=Set2&n=3}}: files have a green color ({\color{graph_node_entity}{\Large $\bullet$}}) and contributors have an orange color ({\color{graph_node_agent}{\Large $\bullet$}}). Edges use colors from the \emph{``3-class Set1''} schema\footnote{\url{https://colorbrewer2.org/?type=qualitative&scheme=Set1&n=3}}: contributions from team members have a blue color ({\color{graph_edge_team}{$\longrightarrow$}}) and contributions from external contributors have a red color ({\color{graph_edge_contributor}{$\longrightarrow$}}). While the chosen colors are `print-friendly', they are not safe regarding color blindness.

The \emph{size of nodes} are proportional to their degree. In our current approach, we generate two drawings for each project; one where we scale the node sizes according to the \emph{in-degree} of file nodes and a second one where we scale according to the \emph{out-degree} of contributors.

For the \emph{layout} we experimented with layout algorithms that are implemented in \textsc{Gephi}, such as \emph{Fruchterman Reingold}~\cite{FruchtermanReingold1991}, the algorithms that comes with \textsc{Graphviz}~\cite{GansnerNorth2000}, and \emph{ForceAtlas2}~\cite{JacomyVenturiniHeymannEtAl2014}.

\begin{table}[h] 
    \caption{Statistics of four selected CWA repositories: \emph{Entities} are the number of files, \emph{Agents} are the number of developers (any role), \emph{Activities} are the number of commits, \emph{Team contributions} are the number of any contribution by CWA team members, \emph{External contr.} are the number of any contribution by external contributors, \emph{Nodes Vis} are the number of nodes in the graph drawing, and \emph{Edges Vis} are the number of edges in the graph drawing.}
    \label{tab:statistics}
    \centering
    \begin{tabular}{lccccccc} 
     \toprule
     \textbf{GitHub Project} & \textbf{Entities} & \textbf{Agents} & \textbf{Activities} & \textbf{Team contr.} & \textbf{Ext. contr.} & \textbf{Nodes Vis} & \textbf{Edges Vis} \\
     \midrule
cwa-server & 4182 & 57 & 366 & 1088 & 849 & 491 & 1209 \\
cwa-documentation & 340 & 31 & 140 & 84 & 45 & 49 & 80 \\
cwa-app-android & 3672 & 56 & 379 & 571 & 1230 & 380 & 1261 \\
cwa-app-ios & 7552 & 53 & 1859 & 809 & 1107 & 287 & 909 \\
     \bottomrule
    \end{tabular}
\end{table}

\begin{figure}[h]
 \centering
 \includegraphics[width=0.7\columnwidth]{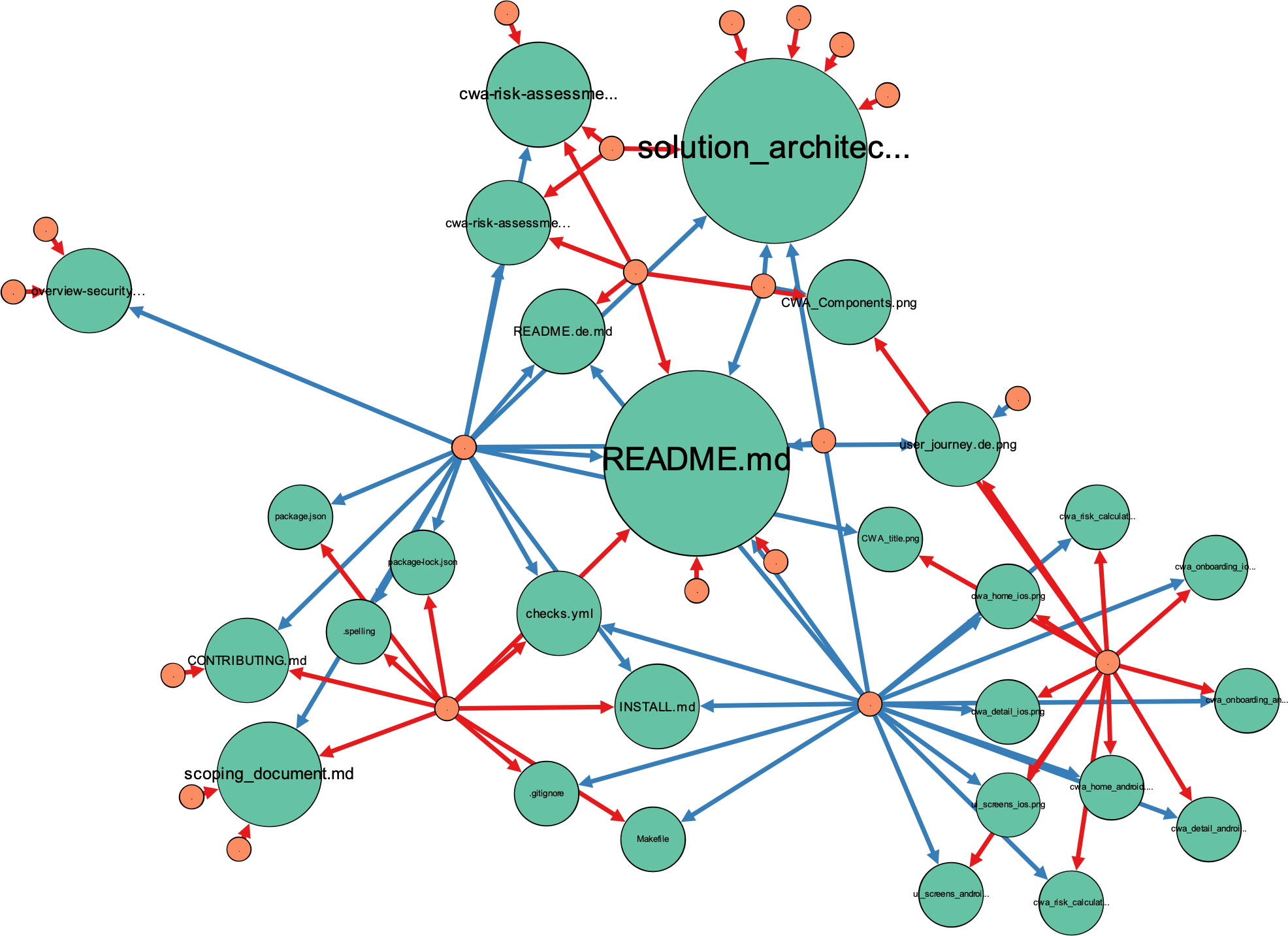}
 \caption{Files ({\color{graph_node_entity}{\Large $\bullet$}}) and contributors ({\color{graph_node_agent}{\Large $\bullet$}}) for the \texttt{cwa-documentation} project.
 Blue edges indicate file changes by team members ({\color{graph_node_agent}{\Large $\bullet$}}{\color{graph_edge_team}{$\longrightarrow$}}{\color{graph_node_entity}{\Large $\bullet$}}). Red edges indicate file changes by external contributors ({\color{graph_node_agent}{\Large $\bullet$}}{\color{graph_edge_contributor}{$\longrightarrow$}}{\color{graph_node_entity}{\Large $\bullet$}}).
 }
 \label{fig:cwa-documentation-entity-in-degree}
\end{figure}

See Figure~\ref{fig:cwa-documentation-entity-in-degree} for an example of a graph drawing for an relatively small project using the \emph{ForceAtlas2} layout algorithm.

\section{Graph Drawings for the Corona-Warn-App}
\label{sec:example_cwa}

The Corona-Warn-App (CWA) has been developed in a short time frame: development started in April 2020 and the app was released on \formatdate{16}{6}{2020} for Android and iOS\@. CWA is developed by SAP and Telekom using an open development process---publicly available from 12 repositories\footnote{\url{https://github.com/corona-warn-app}}. CWA has a decentralized architecture, accompanied by centrally-managed Java-based server applications to distribute findings about infected users and store test results uploaded by the laboratories.

We selected four of the CWA projects for visualization, for which we stored the provenance in \textsc{Neo4j}\footnote{The database dump is available~\cite{Schreiber2020} as of \formatdate{27}{7}{2020}}. These projects differ in their projects statistics regarding number of files in the repository, number of contributing developers, number of commits, and number of files where both team members and external developers made changes---which all leads to different number of nodes and edges for the graph drawings (Table~\ref{tab:statistics}). 

\begin{figure*}[h]
 \centering
 \subfloat[\emph{Entity In-Degree}: Size of nodes according to in-degree \newline of nodes that represent files.]{\label{fig:cwa-app-android-entity-in-degree}
 \includegraphics[width=0.49\columnwidth]{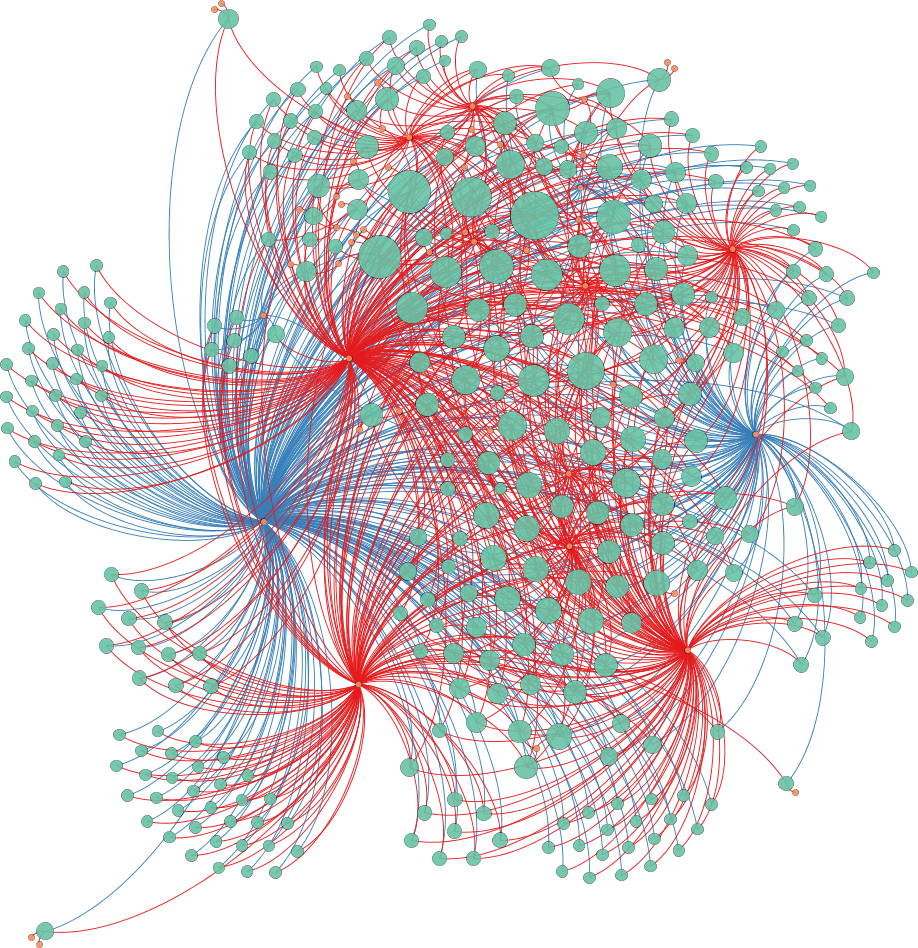}}
 \subfloat[\emph{Agent Out-Degree}: Size of nodes according to out-degree \newline  of nodes that represent contributors.]{\label{fig:cwa-app-android-agent-out-degree}
 \includegraphics[width=0.49\columnwidth]{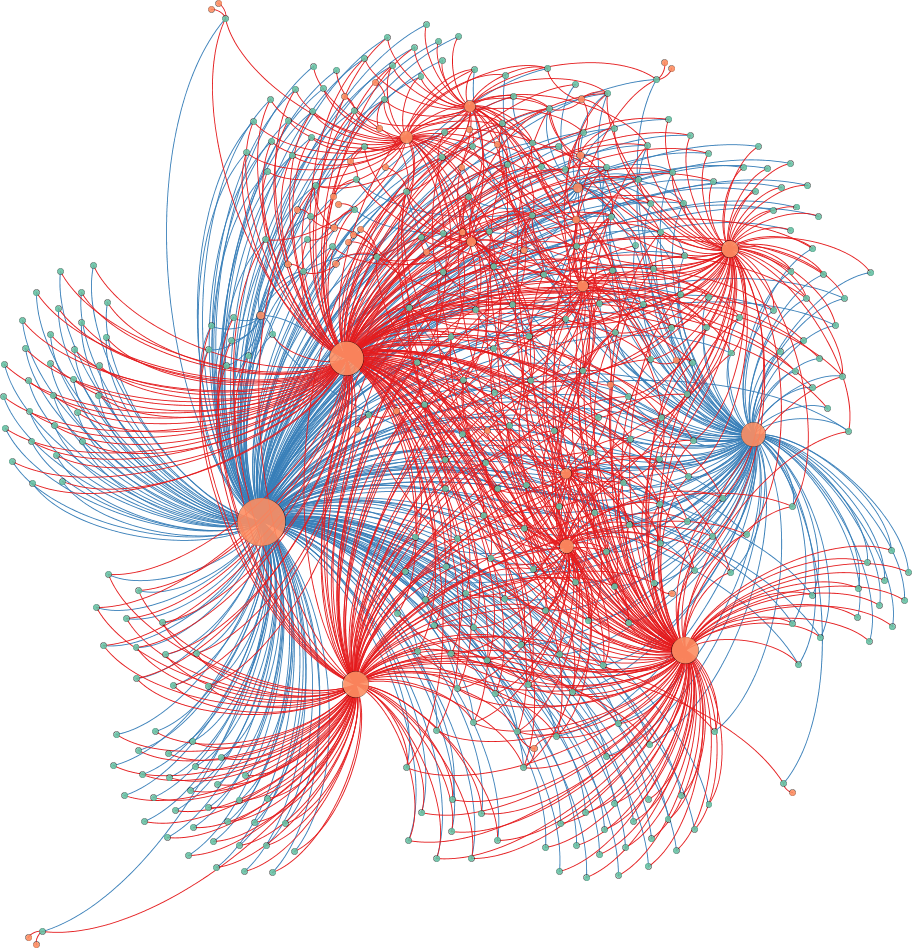}}
 \caption{Files ({\color{graph_node_entity}{\Large $\bullet$}}) and contributors ({\color{graph_node_agent}{\Large $\bullet$}}) for the \texttt{cwa-app-android} project.
 Red edges indicate file changes by team members ({\color{graph_node_agent}{\Large $\bullet$}}{\color{graph_edge_contributor}{$\longrightarrow$}}{\color{graph_node_entity}{\Large $\bullet$}}). Blue edges indicate file changes by external contributors ({\color{graph_node_agent}{\Large $\bullet$}}{\color{graph_edge_team}{$\longrightarrow$}}{\color{graph_node_entity}{\Large $\bullet$}}).
 }
 \label{fig:cwa-app-android}
\end{figure*}

\begin{figure*}[h]
 \centering
 \subfloat[\emph{Entity In-Degree}: Size of nodes according to in-degree  \newline of nodes that represent files.]{\label{fig:cwa-app-ios-entity-in-degree}
 \includegraphics[width=0.49\columnwidth]{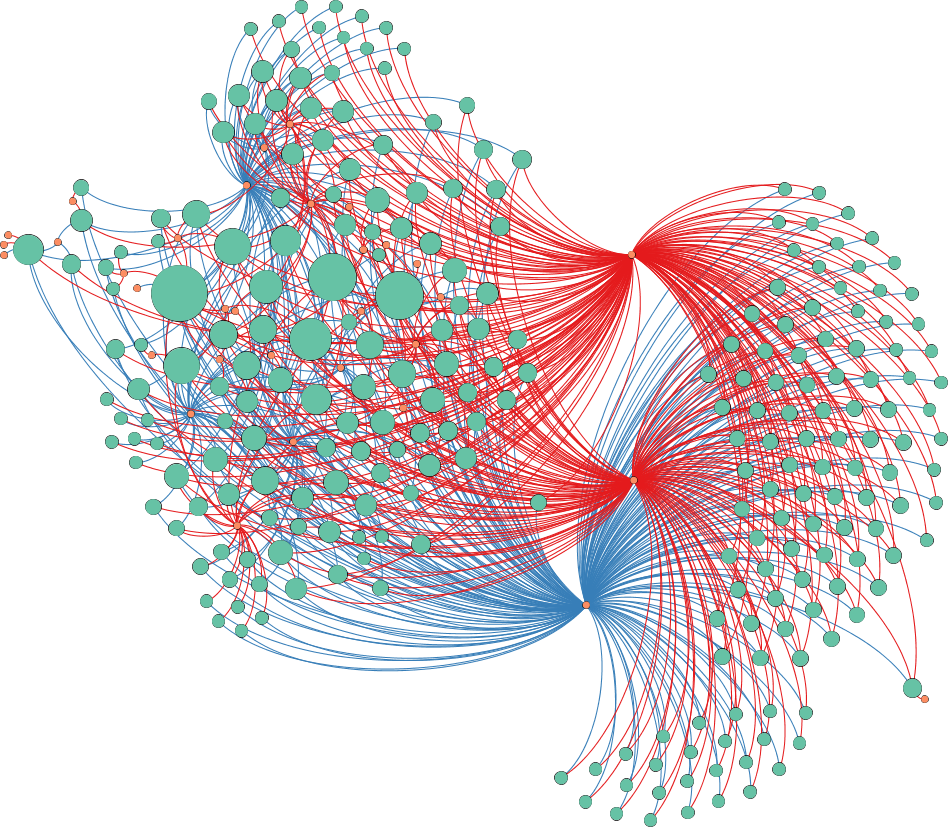}}
 \subfloat[\emph{Agent Out-Degree}: Size of nodes according to out-degree of nodes that represent contributors.]{\label{fig:cwa-app-ios-agent-out-degree}
 \includegraphics[width=0.49\columnwidth]{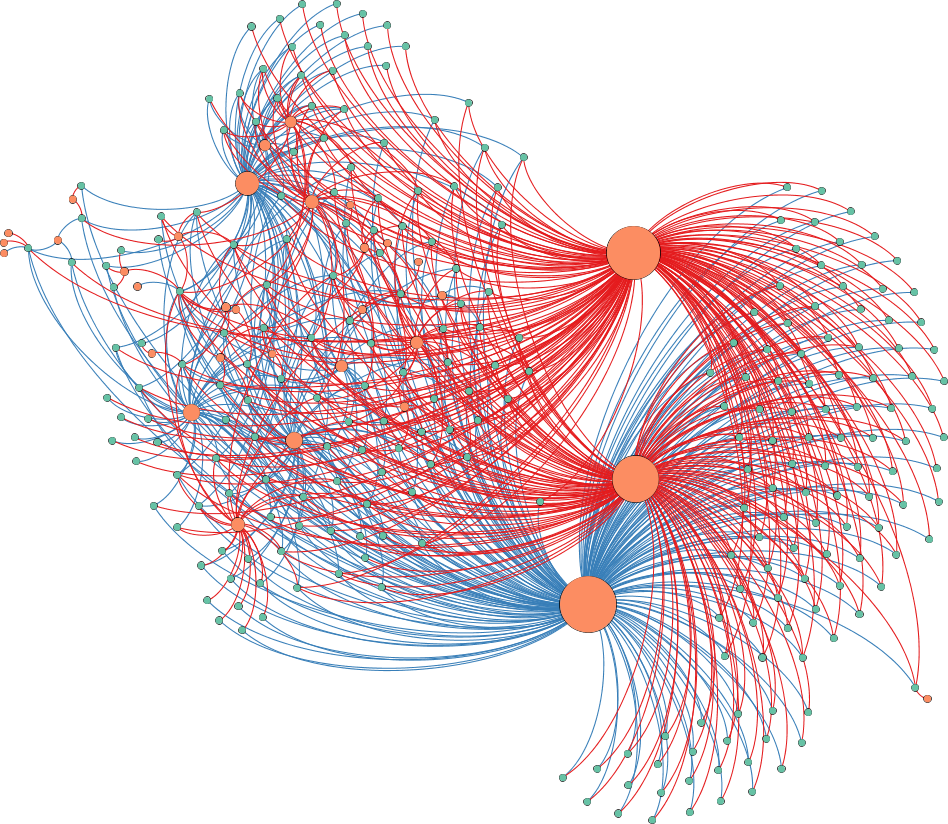}}
 \caption{Files ({\color{graph_node_entity}{\Large $\bullet$}}) and contributors ({\color{graph_node_agent}{\Large $\bullet$}}) for the \texttt{cwa-app-ios} project.
 Red edges indicate file changes by team members ({\color{graph_node_agent}{\Large $\bullet$}}{\color{graph_edge_contributor}{$\longrightarrow$}}{\color{graph_node_entity}{\Large $\bullet$}}). Blue edges indicate file changes by external contributors ({\color{graph_node_agent}{\Large $\bullet$}}{\color{graph_edge_team}{$\longrightarrow$}}{\color{graph_node_entity}{\Large $\bullet$}}).
 }
 \label{fig:cwa-app-ios}
\end{figure*}

For each project, we generate two graph drawings with \textsc{Gephi}\footnote{The \textsc{Gephi} file is available~\cite{Schreiber2020a}.} as described in Section~\ref{sec:graph_visualization}: one where we scale node sizes proportional to the \emph{in-degree} of file nodes (see Figures~\ref{fig:cwa-app-android-entity-in-degree} and~\ref{fig:cwa-app-ios-entity-in-degree}) and a second one where we scale proportional to the \emph{out-degree} of contributors (see Figures~\ref{fig:cwa-app-android-agent-out-degree} and~\ref{fig:cwa-app-ios-agent-out-degree}).

In our graph drawing, typical patterns are visible: team members and external contributors work collaboratively on many files. Because the drawing are based on provenance data, the interpretation is that over the time of development many files were changed by developers with different roles; where a small numbers of developers made most of the changes.

Further, more detailed interpretations and studies of the graph drawing metrics for faithfulness and readability is ongoing work.

\section{Related Work}
\label{sec:related_work}

There are many tools for dynamic history visualization of repository changes over time. A widely used tool is \textsc{Gource}\footnote{\url{https://gource.io}}, which generated movies that show changed files and developer activities. This different to our approach, since we visualize ``condensed'' information about the development history that is stored in the provenance data.

Especially for visualizing social interaction in open-source software projects, Ogawa et al.~\cite{OgawaMaBirdEtAl2007} use an intuitive, time-series, interactive summary view of the social groups that form, evolve and vanish during the entire lifetime of the project.

\section{Conclusion and Future Work}
\label{sec:conclusion}

We presented graph drawings to visually see how team members and external contributor worked on the same files in open-source projects over the course of development.

Since our goal is better understanding of such development patterns, future work foremost is to conduct user studies to evaluate readability and faithfulness. The graph drawings surely can be improved in many ways, for example, with other layouts, color schemes (especially to support color blindness), transparency, or shapes.

We plan, to apply our methods to other projects than CWA; especially, to huge projects with a very long development history. We plan to compare different projects, where the proportion of regular team member and external contributors is different. 

We already work on using the provenance data for non-visual analytics of open-source projects. For example, to investigate whether vulnerabilities are introduced by external contributors (e.g., via pull requests)---we apply static code analysis for revisions in development history determined on the provenance data~\cite{SonnekalbHeinzeKurnatowskiEtAl2020}.

\bibliographystyle{plainurl}  
\bibliography{arXiv-visualization-of-open-source-contributions}  

\end{document}